\documentclass[aps,prl,preprintnumbers,twocolumn,superscriptaddress,nofootinbib,floatfix,10pt]{revtex4-2}
\usepackage{amsfonts,amssymb,stmaryrd,latexsym,amsmath,braket}
\usepackage[caption=false]{subfig}
\usepackage{graphicx,mathtools}
\usepackage{times}
\usepackage{slashed}
\usepackage{braket}
\usepackage{comment}
\usepackage[utf8]{inputenc}
\newcommand{\beginsupplement}{%
        \setcounter{table}{0}
        \renewcommand{\thetable}{S\arabic{table}}%
        \setcounter{figure}{0}
        \renewcommand{\thefigure}{S\arabic{figure}}%
        \setcounter{equation}{0}
        \renewcommand{\theequation}{S\arabic{equation}}        
     }

\begin{document}
        
\title{Floating block method for quantum Monte Carlo simulations}

\author{Avik Sarkar}
\email{a.sarkar@fz-juelich.de}
\affiliation{Institut f\"{u}r Kernphysik, Institute for Advanced Simulation and J\"{u}lich Center for Hadron Physics, Forschungszentrum J\"{u}lich, D-52425 J\"{u}lich, Germany}
\affiliation{Facility for Rare Isotope Beams and Department of Physics and Astronomy,
Michigan State University, East Lansing, MI 48824, USA}

\author{Dean Lee}
\email{leed@frib.msu.edu}
\affiliation{Facility for Rare Isotope Beams and Department of Physics and Astronomy,
Michigan State University, East Lansing, MI 48824, USA}

\author{Ulf-G.~Mei{\ss}ner}
\email{meissner@hiskp.uni-bonn.de}
\affiliation{Helmholtz-Institut f\"{u}r Strahlen- und Kernphysik and Bethe Center for Theoretical Physics, Universit\"{a}t Bonn, D-53115 Bonn, Germany}
\affiliation{Institut f\"{u}r Kernphysik, Institute for Advanced Simulation and J\"{u}lich Center for Hadron Physics, Forschungszentrum J\"{u}lich, D-52425 J\"{u}lich, Germany}
\affiliation{Tbilisi State University, 0186 Tbilisi, Georgia}

        \begin{abstract}
        Quantum Monte Carlo simulations are powerful and versatile tools for the quantum many-body problem. In addition to the usual calculations of energies and eigenstate observables, quantum Monte Carlo simulations can in principle be used to build fast and accurate many-body emulators using eigenvector continuation or design time-dependent Hamiltonians for adiabatic quantum computing.  These new applications require something that is missing from the published literature, an efficient quantum Monte Carlo scheme for computing the inner product of ground state eigenvectors corresponding to different Hamiltonians.  In this work, we introduce an algorithm called the floating block method, which solves the problem by performing Euclidean time evolution with two different Hamiltonians and interleaving the corresponding time blocks.  We use the floating block method and nuclear lattice simulations to build eigenvector continuation emulators for energies of $^4$He, $^8$Be, $^{12}$C, and $^{16}$O nuclei over a range of local and non-local interaction couplings.  From the emulator data, we identify the quantum phase transition line from a Bose gas of alpha particles to a nuclear liquid.
        \end{abstract}
\maketitle

\section{Introduction}
Quantum Monte Carlo simulations are widely used for first-principles calculations of solid state systems and condensed matter 
\cite{Zhang:1997, foulkes2001quantum, Drut:2009aj, kolorenvc2011applications, Ostmeyer:2020uov, Luu:2022efw},
quantum chemistry \cite{umrigar1993diffusion, hammond1994monte, morales2012multideterminant}, atomic and molecular physics \cite{lester1990quantum, booth2010approaching, morales2012multideterminant}, lattice field theories \cite{creutz1983monte, degrand2006lattice, gattringer2009quantum}, nuclear physics \cite{Dean:2009, Carlson:2014vla, Gezerlis:2014, lahde2019nuclear}, degenerate quantum gases \cite{Ceperley:1980, astrakharchik2004equation, capogrosso2008monte}, and other quantum many-body systems.  In cases where sign oscillations are under control, the computational effort scales favorably as a low-order polynomial in the number of particles.  There are many examples of quantum many-body systems with strong correlations where quantum Monte Carlo simulations are the only tools currently available for reliable and accurate predictions.  The standard quantities calculated in quantum Monte Carlo simulations are low-lying energy levels and their corresponding eigenstate observables.  Since wave functions are not constructed explicitly, the set of linear algebra operations that can be performed on energy eigenstates is often more limited than that for other methods that explicitly construct and store wave functions.  

One basic linear algebra operation that would be extremely useful is the calculation of inner products between energy eigenstates corresponding with different quantum Hamiltonians.  Calculating eigenstate inner products within the framework of quantum Monte Carlo simulations would allow for the construction of fast and accurate emulators using eigenvector continuation \cite{Frame:2017fah} for systems that might otherwise be inaccessible using other methods.  Another potential application is the ability to use quantum Monte Carlo simulations to design time-dependent Hamiltonians $H(t)$ for efficient adiabatic quantum computing \cite{farhi2000quantum, farhi2001quantum}. The initial Hamiltonian $H(0)$ is any trivial Hamiltonian whose ground state can be prepared on a quantum computer, and the final Hamiltonian $H(T)$ is the quantum Hamiltonian of interest.  Starting from the ground state of $H(0)$ and slowly evolving with $H(t)$, one can accurately prepare the ground state of $H(T)$ when $T$ is sufficiently large.  Quantum Monte Carlo simulations on classical computers can be used to optimally select $H(0)$ and the time-dependent path $H(t)$ such that the inner products between ground states at times $t$ and $t+dt$ remain as large as possible for each $t$, thereby reducing the need for very slow time evolution.  These calculations require calculations of inner products between the ground states corresponding to $H(t)$ and $H(t+dt)$.  This information can also be used to compute Berry connections, curvatures, and phases associated with cyclical adiabatic evolution \cite{Berry:1984jv}. After preparing the ground state of the desired Hamiltonian $H(T)$, the quantum computer could then be used to perform real-time dynamics that would otherwise be beyond classical computing capabilities.   

In this work, we introduce a quantum Monte Carlo algorithm called the floating block method that computes the inner product between two eigenstates produced by two different Hamiltonians.  This is achieved by performing Euclidean time evolution for the two Hamiltonians and interleaving the corresponding time blocks.  We demonstrate the floating block method using nuclear lattice simulations and eigenvector continuation to build emulators for the energies of $^4$He, $^8$Be, $^{12}$C, and $^{16}$O nuclei over a range of local and non-local interaction couplings.  

\section{Floating block method}
Let $H_i$ and $H_j$ be two different quantum Hamiltonians with ground state wave functions $\ket{v^0_i}$ and $\ket{v^0_j}$, respectively, and ground state energies $E^0_i$ and $E^0_j$, respectively.  The only requirement is that $H_i$ and $H_j$ can be simulated using quantum Monte Carlo simulations.  The floating block method is based on the identity,
\begin{equation}
\lim_{t \rightarrow \infty}\frac{\braket{\psi_I|e^{-H_i t}e^{-H_j t}e^{-H_i t}e^{-H_j t}|\psi_I}}{\braket{\psi_I|e^{-2H_i t}e^{-2H_j t}|\psi_I}}=|\braket{v^0_i|v^0_j}|^2, \label{FBequation}
\end{equation}
where $\ket{\psi_I}$ is any initial state that is not orthogonal to both $\ket{v^0_i}$ and $\ket{v^0_j}$. We note that the ground state energy values $E^0_i$ and $E^0_j$ drop out of Eq.~\eqref{FBequation} in the limit of large~$t$.  This is a key feature and computational advantage of the floating block method.  For cases where the phase of the inner product is nontrivial, we calculate the phase using
\begin{equation}
\lim_{t \rightarrow \infty}\frac{\braket{\psi_I|e^{-2H_i t}e^{-2H_j t}|\psi_I}}{|\braket{\psi_I|e^{-2H_i t}e^{-2H_j t}|\psi_I}|}=\frac{\braket{v^0_i|v^0_j}}{|\braket{v^0_i|v^0_j}|}. \label{phase}
\end{equation}  
We are using the phase convention that $\braket{\psi_I|v^0_i}$ and $\braket{\psi_I|v^0_j}$ are positive.
The floating block method can be used for quantum Monte Carlo simulations on the lattice or in continuous space and is compatible with path integral Monte Carlo simulations where particle worldlines are explicitly sampled or auxiliary-field Monte Carlo simulations where the particles are integrated out.

One can also compute $\braket{v^0_i|v^0_j}$ using the reweighting techniques used in Ref.~\cite{Frame:2017fah, Frame:2019jsw}.  This approach is presented in the Supplemental Material.  Unfortunately, the reweighting approach is viable only when the system is small in size, the Euclidean time $t$ is not too large, and the ground state energies for the different Hamiltonians are numerically close.  We show in the Supplemental Material that the floating block method provides a computational advantage over reweighting calculations equal to several orders of magnitude.

\section{Applications to eigenvector continuation}
Eigenvector continuation (EC) is a variational technique for finding the extremal eigenvectors and eigenvalues of a parameter-dependent Hamiltonian matrix $H(c)$ \cite{Frame:2017fah}. The method relies on the analyticity of the eigenvectors as a function of parameters of the Hamiltonian.  The resulting smooth parametric dependence allows the eigenvector manifold to be well-approximated by a low-dimensional subspace. By reducing the dimensionality of the problem, calculations are orders of magnitude faster, and this aspect has been used to build fast and accurate emulators using EC \cite{Konig:2019adq, Ekstrom:2019lss}. Its convergence properties were investigated in \cite{Sarkar:2021sal}, and a greedy algorithm to optimally select training points was presented in Ref.~\cite{Sarkar:2022}.  In Ref.~\cite{Bonilla:2022rph, Melendez:2022kid}, EC was identified as a specific example of a wider class of techniques called reduced basis methods \cite{Reduced_basis_book, Quarteroni_2011, Field_2011}.   Meanwhile, there have been numerous theoretical developments and applications of eigenvector continuation and other reduced basis methods in nuclear physics \cite{Demol:2019yjt, Frame:2019jsw, Furnstahl:2020abp, Bai:2021xok, Wesolowski:2021cni, Yoshida:2021jbl, Melendez:2021lyq, Melendez:2022kid}.

Let us consider a Hamiltonian of the form $H(c) = H_0 + cH_1$.  In order to perform EC calculation of the ground state, we first choose some set training points $c=c_0, \cdots, c_N$ and find the exact ground state eigenvectors $\ket{v^0_0}, \cdots, \ket{v^0_N}$. If our target system corresponds to parameter value $c=c_t$, we calculate the projected Hamiltonian matrix $H_{ij} = \braket{v^0_i|H(c_t)|v^0_j}$, norm matrix $N_{ij} = \braket{v^0_i|v^0_j}$, and solve the corresponding generalized eigenvalue problem.  We use the floating block method to compute the norm matrix $N_{ij}$.  In order to determine the projected Hamiltonian matrix elements, we use the result
\begin{equation}
\lim_{t \rightarrow \infty}\frac{\braket{\psi_I|e^{-2H_i t}H(c_t)e^{-2H_j t}|\psi_I}}{\braket{\psi_I|e^{-2H_i t}e^{-2H_j t}|\psi_I}}=\frac{H_{ij}}{N_{ij}}. \label{Eq:H_over_N}
\end{equation}

\section{Methods}
Reviews of auxiliary-field lattice Monte Carlo simulations are given in Ref.~\cite{Dean:2009,lahde2019nuclear}. We start with an initial state and propagate it over a large but finite number of time steps, which we denote by $L_t$. In each of these time steps, we multiply the current state by the transfer matrix $M = \exp(-H(c)\Delta t)$.  For notational simplicity, we omit the normal ordering symbols $::$ that should enclose each transfer matrix. 
 From the data we obtain from different $L_t$, we can extrapolate to infinite time. 

Consider a family of Hamiltonians of the form $H(c) = H_0 + cH_1$.  For the lattice simulations presented here, we consider two independent parameters, but the formalism is the same.  We are interested in performing calculations at some target value point $c = c_t$, using the training data at $c = \{c_1,\cdots,c_k\}$.  
For a finite number of time steps, $L_t$, we write Eq.~\eqref{Eq:H_over_N} as
\begin{align}
    \frac{H_{ij}}{N_{ij}} = \frac{\braket{\psi_{I}|e^{-H(c_i)\Delta t}|\cdots|H(c_t)|\cdots|e^{-H(c_j)\Delta t}|\psi_{I}}}{\braket{\psi_{I}|e^{-H(c_i)\Delta t}|\cdots|e^{-H(c_j)\Delta t}|\psi_{I}}}.
\end{align}
For notational convenience, the limit $L_t \rightarrow \infty$ on the right-hand side is implicit.  We further simplify the notation as
\begin{align}
    \frac{H_{ij}}{N_{ij}} = \frac{\braket{\psi_{I}|c_i|\cdots|c_i|H(c_t)|c_j|\cdots|c_j|\psi_{I}}}{\braket{\psi_{I}|c_i|\cdots|c_i|c_j|\cdots|c_j|\psi_{I}}},
\end{align}
where $|c_i|$ represents a time step where we multiply by transfer matrix $e^{-H(c_i)\Delta t}$, and there are $L_t/2$ time steps of $|c_i|$ and $|c_j|$ each. Note again that since there are an equal number of such time steps in the numerator and denominator, the ground state energy factors cancel out in the limit $L_t \rightarrow \infty$.
The squared magnitude of the norm matrix $N_{ij}$ is given by
\begin{align}
     |N_{ij}|^2 = \frac{\braket{\psi_{I}|c_i|\cdots|c_i|c_j|\cdots|c_j|c_i|\cdots|c_i|c_j|\cdots|c_j|\psi_{I}}} {\braket{\psi_{I}|c_i|\cdots|c_i|c_i|\cdots|c_i|c_j|\cdots|c_j|c_j|\cdots|c_j|\psi_{I}}}\label{N_ij}.
\end{align}
Again, the limit $L_t \rightarrow \infty$ is implicit.  We prove these results in the Supplemental Material.

For our quantum Monte Carlo simulations, we select configurations according to the absolute value of the expression in the denominator of Eq.~\eqref{N_ij} and evaluate the expressions in the numerator and the denominator. However, the stochastic noise will be large if the expressions in the numerator and denominator are very different. We first discuss how to optimize the performance of the floating block method for auxiliary-field Monte Carlo simulations, and then we discuss how to optimize the process for path integral Monte Carlo simulations without auxiliary fields.  

We can reduce the stochastic error in auxiliary-field Monte Carlo simulations with two techniques.  The first is to reorder the auxiliary fields in the numerator. Usually, the same auxiliary field configurations are used for calculating the numerator and the denominator in Eq.~\eqref{N_ij}.  Here, we instead reorder the auxiliary fields in the numerator so that the sequence of auxiliary fields for coupling $c_i$ is the same for the numerator and denominator.  Similarly, the sequence of auxiliary fields for coupling $c_j$ is the same for the numerator and denominator.  After performing this reordering, the only difference between the numerator and denominator comes from the commutator of the transfer matrices with different Hamiltonian couplings.  We can illustrate the calculation of $|N_{ij}|^2$ with 12 time steps in the top line of Fig.~\ref{N_ij}.  The time steps in the numerator and denominator connected by double-arrow lines have the same auxiliary field configurations. This reordering greatly reduces the noise of the Monte Carlo simulations. Without the reordering, it is impossible to compute norm matrices from any calculation with more than a few time steps. We discuss the computational advantage of reordering the auxiliary fields in the Supplemental Material and show that it provides a computational advantage of many orders of magnitude.

\begin{figure}[hbt!]
\includegraphics[width=8.4cm]{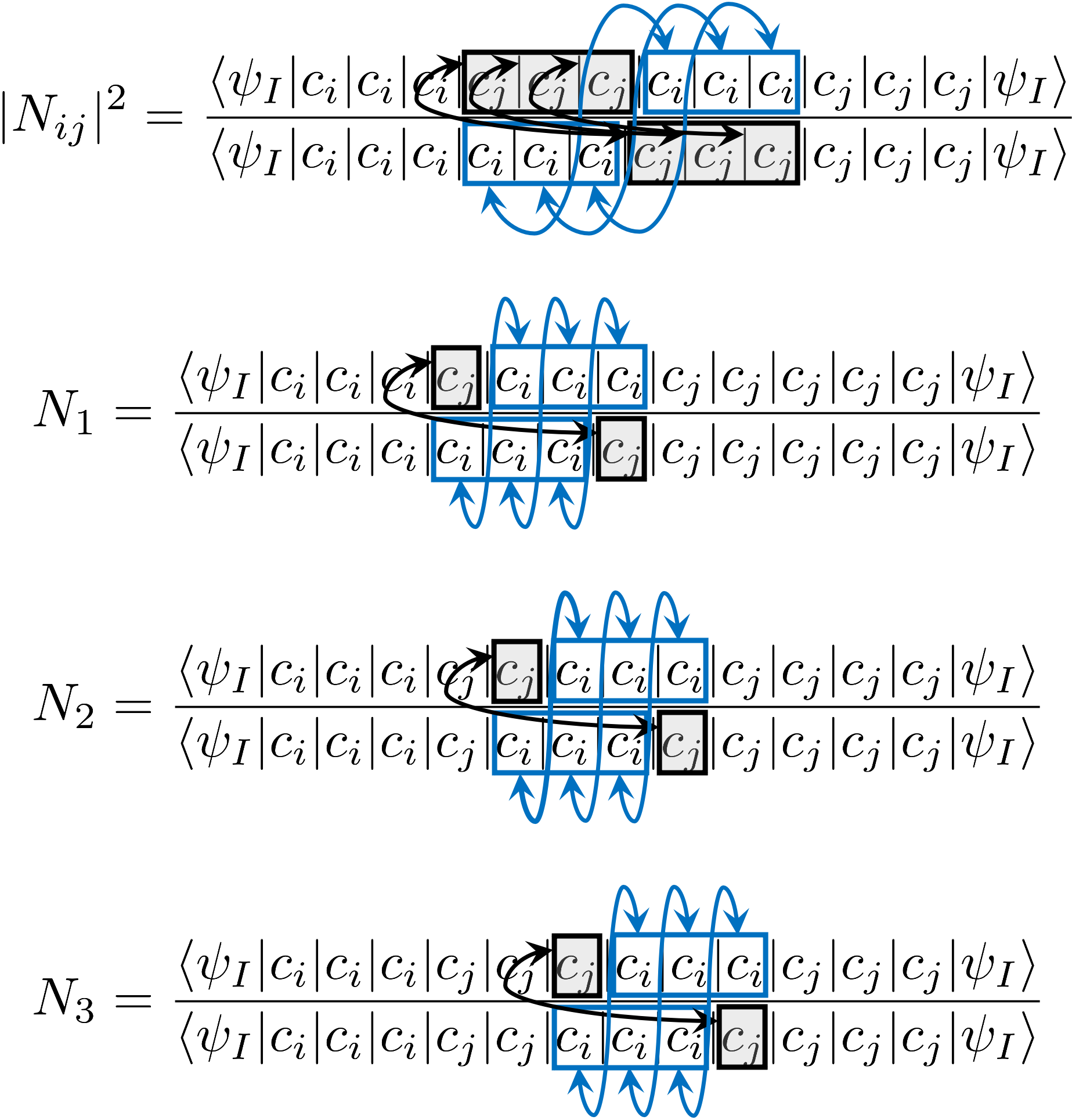}
\caption{The norm matrix calculation of $|N_{ij}|^2$ with 12 time steps.}
\end{figure}

When the displacement of time blocks is large, the numerator and denominator are quite different calculations, and the stochastic noise can become quite large.  This is fixed by the second of our two techniques.  We allow the time blocks to ``float'' gradually, similar to a block of ice detaching from a large ice mass and floating into the sea.  In the following example, we float three time blocks one step at a time.  This is illustrated in the bottom three lines of Fig.~\ref{N_ij}.  Each term, $N_1, N_2, N_3$, is computed separately and multiplied together to give $N_1 N_2 N_3 = |N_{ij}|^2$.  This gradual floating of time blocks is also useful for path integral Monte Carlo simulations without auxiliary fields.   For path integral Monte Carlo simulations, we work with particle worldlines defined by particle positions as functions of Euclidean time.  We again use gradual floating of the time blocks to produce correlated samples with reduced stochastic noise.  For each ratio of amplitudes we calculate, the worldlines are chosen to be the same for the numerator and denominator.

\section{Results}

We perform nuclear lattice simulations using a simple leading order interaction that is independent of spin and isospin.  Despite the simplicity of the interaction, previous lattice studies have shown that they produce a good description of the bulk properties of nuclear structure and thermodynamics \cite{Lu:2018bat, Lu:2019nbg, Shen:2022bak, Ren:2023ued}.  We use a spatial lattice spacing of 1.97 fm and a time lattice spacing of 0.197 fm$/c$.  The interactions are composed of two types of interactions, which we write as $V_L({\bf r}',{\bf r})$ and $V_{NL}({\bf r}',{\bf r})$.  Here ${\bf r}$ and ${\bf r}'$ are the relative separation of the incoming and outgoing nucleons respectively. For $V_{NL}$ the interaction is non-local, which means that ${\bf r}$ and ${\bf r}'$ are in general different. Meanwhile, $V_L$ is a local interaction where ${\bf r}'$ and ${\bf r}$ are the same. These local and non-local interactions are smeared in space with parameters $s_L$ and $s_{NL}$ respectively, and the normalizations of $V_L$ and $V_{NL}$ are chosen so that the physical ${}^{4}$He binding energy is reproduced for interactions $H_A = K + V_L$ and $H_B = K + V_{NL}$, where $K$ is the kinetic energy operator.  We describe the interactions in detail in the Supplemental Material. 

We consider general linear combinations of the local and non-local interactions, $H(c_L,c_{NL}) = K + c_LV_L + c_{NL}V_{NL}$, and compute the binding energies of ${}^{4}$He, ${}^{8}$Be, ${}^{12}$C, and ${}^{16}$O.  In Table~\ref{Table:exact_error}, we show EC predictions for the ground state energy of ${}^{12}$C using the floating block method in comparison with full simulation results, for $L_t = 400$. For the EC emulator, we use the training points $(c_L,c_{NL}) = (0.5,0.5)$, $(0.2,0.8)$, $(0,1)$ and $(0.2,0.6)$, with the corresponding ground state vectors included in that order.  The reported errors for the full simulation are one standard deviation stochastic errors.  The errors for EC calculations are also one standard deviation errors determined from propagating stochastic errors using the trimmed sampling method described in Ref.~\cite{Caleb:2023}.  We see that the EC emulator very accurately reproduces the full simulation results for test points $(c_L,c_{NL})$ within the interpolation region of the training points.  For test points outside of the interpolation region, the quality of the predictions is less accurate but still quite good for most cases. 

We note that the EC emulator is a variational approximation.  The error bars shown at each EC order in Table~\ref{Table:exact_error} correspond to the uncertainty in the variational approximation predictions.  As more orders in the EC approximation are used, the EC results should converge to the full simulation results from above.


\begin{table*}[hbt!]
\centering{}
\begin{tabular}{|c | c| c| c| c|}
\hline
 $(c_L,c_{NL})$ &Full Simulation & 2nd order EC & 3rd order EC & 4th order EC\\ 
 \hline
 (0.8,0.2) & $-338.57 \pm 0.03$ & $-330.63 \pm 0.26$ & $-333.15 \pm1.85$ & $-333.24 \pm 1.14$\\ 
 (0.8,0.1) & $-295.33 \pm 0.02$ & $-290.04 \pm 0.24$ & $-292.08 \pm 1.51$ & $-292.17 \pm 1.03$\\ 
 (0.9,0.1) & $-381.81 \pm 0.02$ & $-369.12 \pm 0.28$ & $-372.63 \pm 2.24$ & $-372.77 \pm 1.83$\\ 
 (0.8,0.3) & $-382.41 \pm 0.03$ & $-371.22 \pm 0.27$ & $-374.24 \pm 2.25$ & $-374.34 \pm 1.42$\\
 (0.4,0.6) & $-177.15 \pm 0.05$ & $-177.25 \pm 0.19$ & $-177.32 \pm 0.46$ & $-177.33 \pm 0.24$\\ 
 (0.4,0.7) & $-217.73 \pm 0.04$ & $-217.70 \pm 0.19$ & $-217.71 \pm 0.38$ & $-217.72 \pm 0.21$\\
 (0.3,0.7) & $-141.11 \pm 0.05$ & $-139.63 \pm 0.19$ & $-141.09 \pm 0.71$ & $-141.10 \pm 0.35$\\
 (0.4,0.8) & $-259.35 \pm 0.06$ & $-258.19 \pm 0.20$ & $-258.31 \pm 0.30$ & $-258.32 \pm 0.28$\\
 (0.2,0.6) & $-41.54 \pm 0.05$  & $-31.91 \pm 0.15$  & $-37.75  \pm 0.50$ & $-41.64 \pm 0.14$\\
 (0.2,0.2) &  $31.26 \pm 0.02$  & $101.48 \pm 0.17$  &  $99.18  \pm 1.86$ &  $69.72 \pm 1.77$\\
 (0.8,0.8) & $-606.63 \pm 0.05$ & $-574.23 \pm 0.37$ & $-579.97 \pm 4.02$ & $-580.12 \pm 2.70$\\
 \hline
\end{tabular}
\caption{Comparison of the EC emulator prediction and full simulations results for ${}^{12}$C. The training points added to the EC subspace progressively are: $(c_L,c_{NL}) = (0.5,0.5)$, $(0.2,0.8)$, $(0,1)$, and $(0.2,0.6)$.
}
\label{Table:exact_error}
\end{table*}

\begin{figure}[hbt!]
    \centering
    (a) $E({}^{8}\text{Be}) - 2E({}^{4}\text{He})$ \\   
    \includegraphics[width=8cm]{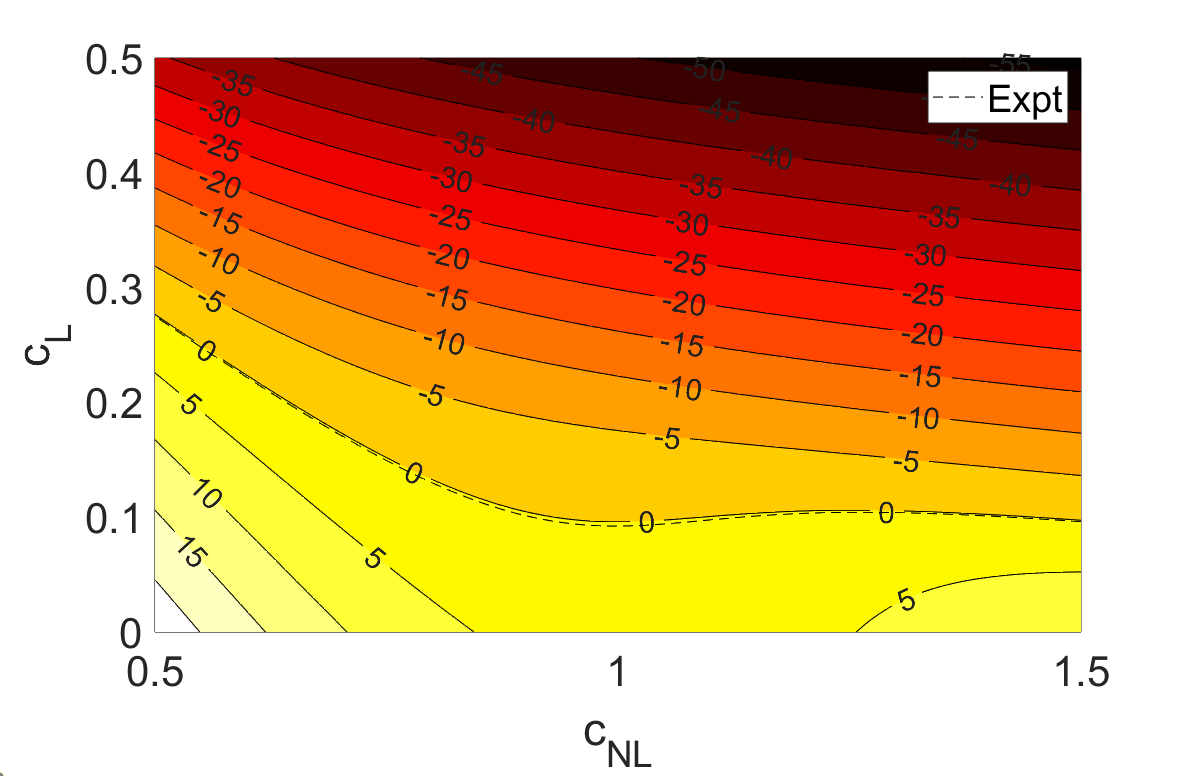} \\
    (b) $E({}^{12}\text{C}) - 3E({}^{4}\text{He})$ \\   
    \includegraphics[width=8cm]{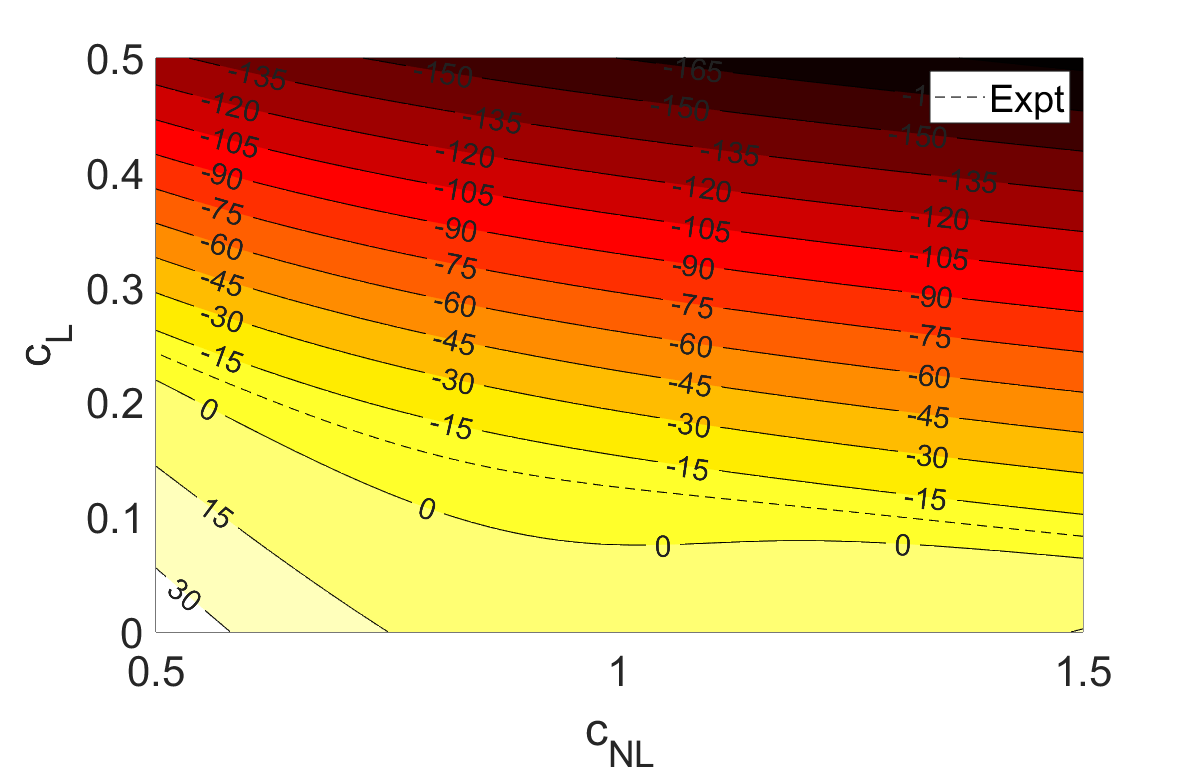} \\
    (c) $E({}^{16}\text{O}) - 4E({}^{4}\text{He})$ \\ 
    \includegraphics[width=8cm]{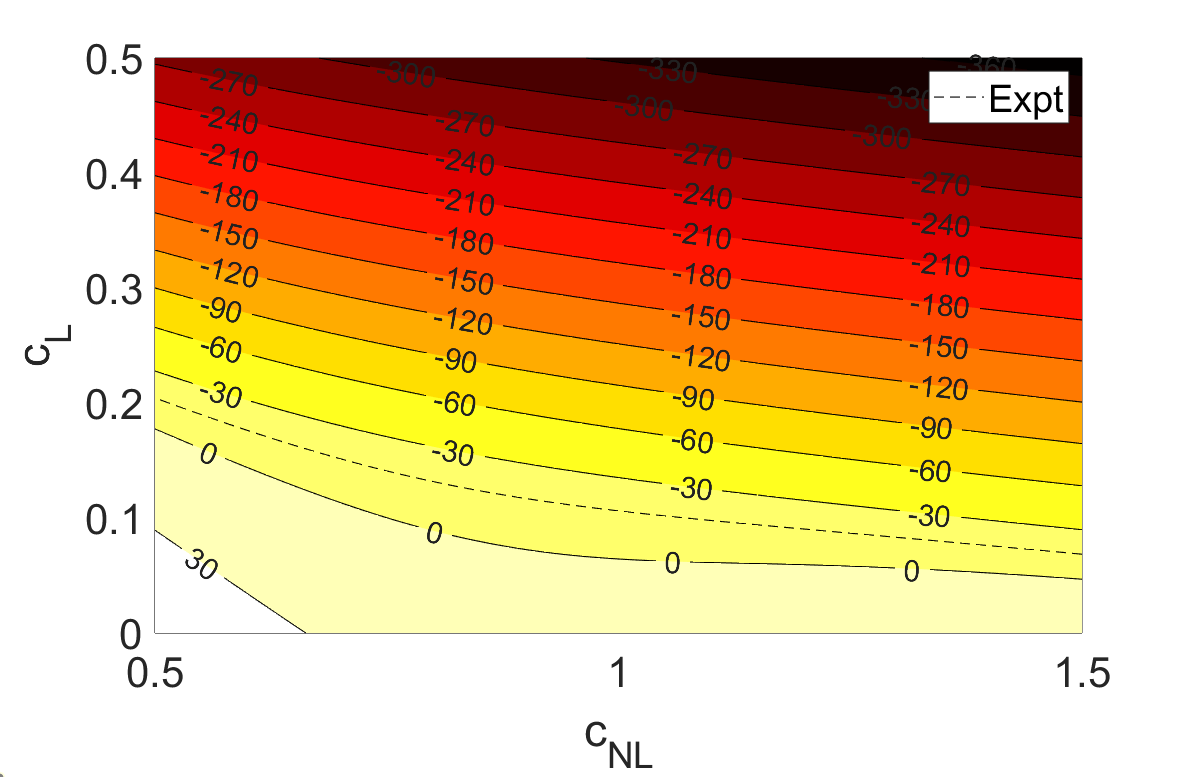} \\ 
\caption{Contour plots for the difference between the EC emulated energy and the corresponding multi-alpha threshold energies for ${}^{8}$Be in Panel (a), ${}^{12}$C in Panel (b), and ${}^{16}$O in Panel (c). The dashed lines show the experimental values.
}
\label{fig:Emulated_contour_plots}
\end{figure}

In Fig.~\ref{fig:Emulated_contour_plots}, we plot the ground state energy of ${}^{8}$Be relative to the two-alpha threshold in Panel (a), ground state energy of ${}^{12}$C relative to the three-alpha threshold in Panel (b), and ground state energy of ${}^{16}$O relative to the four-alpha threshold in Panel (c).  For each case, we perform second-order EC with training points at $(c_L,c_{NL}) = (0.5,0.5)$ and $(0,1)$ in a periodic box of length $L = 15.76$~fm. The dashed lines show the experimentally observed values.  The location of the quantum phase transition in the $(c_L, c_{NL})$ plane can be seen clearly from the zero contours in Fig.~\ref{fig:Emulated_contour_plots} where the ground state energy equals the corresponding multi-alpha threshold.  We note that the quantum phase transition occurs at approximately the same locations for ${}^8$Be, ${}^{12}$C, and ${}^{16}$O, corresponding with the line where the alpha-alpha interaction changes from repulsive to attractive. Consistent with the findings in Ref.~\cite{Serdar:2016}, we see that the local interaction coupling $c_L$ plays a dominant role in determining whether we have a Bose gas or a nuclear liquid.  We cannot produce a nuclear liquid using $c_{NL}$ alone.

\section{Summary and Discussion}

We have introduced the floating block method, a new computational algorithm for quantum Monte Carlo simulations that allows for the calculation of inner products between ground-state wave functions corresponding to different Hamiltonians. 
Such calculations were not practical using previously existing methods, and while they were possible for small systems, even for those systems the floating block method provides a computational advantage of several orders of magnitude that scales with the complexity of the calculation.

This ability to compute inner products between ground-state wave functions corresponding to different Hamiltonians opens the door to efficient construction of EC emulators using quantum Monte Carlo simulations.  In this work, we have used this method to explore the $(c_L, c_{NL})$ phase diagram corresponding to local and non-local interactions.  We find that the EC emulators are very accurate for interpolation and also fairly reliable for extrapolation.  We are able to locate the quantum phase transition line between a Bose gas of alpha particles and a nuclear liquid.

While the examples we have considered here are lattice calculations using auxiliary fields, the application of the floating block method for continuum calculations with path integral Monte Carlo is straightforward.  We hope that this theoretical development will enable the construction of new emulators using quantum Monte Carlo methods as well as possible novel applications for quantum computing.

\section{Acknowledgement}
We are grateful for useful discussions with members of the Nuclear Lattice Effective Field Theory Collaboration.
DL was supported in part by the U.S. Department of Energy (DE-SC0013365, DE-SC0021152, DE-SC0023658, DE-SC0024586, SciDAC-5 NUCLEI Collaboration) and the U.S. National Science Foundation (PHY-2310620).  UGM was supported by the European Research Council (ERC) under the European Union's Horizon 2020 research and innovation programme (grant agreement No. 101018170), by Deutsche Forschungsgemeinschaft (DFG, German Research Foundation) (project-ID 196253076 - TRR 110), the Chinese Academy of Sciences (CAS) President's International Fellowship Initiative (PIFI) (Grant No. 2018DM0034), Volkswagen Stiftung (Grant No. 93562) and by the MKW NRW (funding code NW21-024-A). Computational resources provided by the Oak Ridge Leadership Computing Facility through the INCITE award
``Ab-initio nuclear structure and nuclear reactions'', the Southern Nuclear Science Computing Center, the Gauss Centre for Supercomputing e.V. (www.gauss-centre.eu) for computing time on the GCS Supercomputer JUWELS at the J{\"u}lich Supercomputing Centre (JSC), and the Institute for Cyber-Enabled Research at Michigan State University.

\bibliography{References}
\bibliographystyle{apsrev}
       
\beginsupplement
     
 \clearpage     

\onecolumngrid        

\section{Supplemental Material}

\subsection{Derivation of the floating block method}
Let us consider the following ratio of amplitudes,
\begin{align}
R_{ij} = \frac{\braket{\psi_{I}|c_i|\cdots|c_i|c_j|\cdots|c_j|c_i|\cdots|c_i|c_j|\cdots|c_j|\psi_{I}}} {\braket{\psi_{I}|c_i|\cdots|c_i|c_i|\cdots|c_i|c_j|\cdots|c_j|c_j|\cdots|c_j|\psi_{I}}},
    \label{Eq:Norm_matrix_repeated}
\end{align}
where we use the notation $|c_i|$ for the transfer matrix with coupling $c_i$ and $|c_j|$ for the transfer matrix with coupling $c_j$.
In the numerator, we have four blocks of $L_t/4$ time steps with couplings $|c_i|$ and $|c_j|$ interleaved with each other.  We now project onto eigenstates $\sum_n \ket{v^n_i}\bra{v^n_i}$ for $H(c_i)$ and $\sum_n \ket{v^n_j}\bra{v^n_j}$ for $H(c_j)$.  We then get
\begin{align}
R_{ij} = \frac{\braket{\psi_{I}|c_i|\cdots|c_i|\sum_{n} \ket{v^{n}_i}\bra{v^n_i}|c_j|\cdots|c_j| \sum_{n} \ket{v^{n}_j}\bra{v^n_j}|c_i|\cdots|c_i| \sum_n \ket{v^{n}_i}\bra{v^n_i} c_j|\cdots|c_j| \sum_{n} \ket{v^{n}_j}\bra{v^n_j}\psi_{I}}} {\braket{\psi_{I}|c_i|\cdots|c_i| \sum_{n} \ket{v^{n}_i}\bra{v^n_i}
c_j|\cdots|c_j|
\sum_n \ket{v^{n}_j}\bra{v^n_j} \psi_{I}}}.
\end{align}
If we now take the limit as $L_t$ goes to infinity, we then have
\begin{align}
  R_{ij} = \frac{\braket{\psi_{I}|v^0_i}\braket{v^0_i|v^0_j}\braket{v^0_j|v^0_i}\braket{v^0_i|v^0_j}\braket{v^0_j|\psi_{I}}}
    {\braket{\psi_{I}|v^0_i}\braket{v^0_i|v^0_j} \braket{v^0_j|\psi_{I}}}  = |\braket{v^0_i|v^0_j}|^2 = |N_{ij}|^2,
\end{align}
where $\ket{v^0_i}$ is the ground state eigenvector for $H(c_i)$ and $\ket{v^0_j}$ is the ground state eigenvector for $H(c_j)$.  

\subsection{Lattice interactions}
For our lattice calculations, we use a spatial lattice spacing $a = (100\; \text{MeV})^{-1} = 1.97$~fm, and time lattice spacing $a_t = (1000\;\text{MeV})^{-1} = 0.197$~fm. We take the nucleon mass to be $m=938.92$~MeV.  We use $\sum_{\braket{\mathbf{n'}\mathbf{n}}}$ to denote the sum over nearest-neighbor lattice sites of $\mathbf{n}$. Similarly, $\sum_{\braket{\mathbf{n'}\mathbf{n}}_i}$ denotes sum over nearest-neighbor lattice sites of $\mathbf{n}$ along the $i^{th}$ axis, $\sum_{\braket{\braket{\mathbf{n'}\mathbf{n}}}_i}$ denotes sum over next-to-nearest-neighbor lattice sites of $\mathbf{n}$ along the $i^{th}$ axis, and $\sum_{\braket{\braket{\braket{\mathbf{n'}\mathbf{n}}}}_i}$ denotes sum over next-to-next-to-nearest-neighbor lattice sites of $\mathbf{n}$ along the $i^{th}$ axis.

For each lattice site $\mathbf{n}$ and smearing parameter $s_{NL}$, we define non-local annihilation and creation operators,
\begin{align}
    a_{NL}(\mathbf{n}) &= a(\mathbf{n}) + s_{NL} \sum_{\braket{\mathbf{n'}\mathbf{n}}} a(\mathbf{n'}),\\
    a_{NL}^\dagger(\mathbf{n}) &= a^\dagger(\mathbf{n}) + s_{NL} \sum_{\braket{\mathbf{n'}\mathbf{n}}} a^\dagger(\mathbf{n'}),
\end{align}
Similar to the density operator $\rho(\mathbf{n}) = a^\dagger(\mathbf{n}) a(\mathbf{n})$, we define the smeared non-local density,
\begin{align}
    \rho_{NL}(\mathbf{n}) = a_{NL}^\dagger(\mathbf{n}) a_{NL}(\mathbf{n}).
\end{align}
The smeared local density is defined as
\begin{align}
    \rho_L(\mathbf{n}) = a^\dagger(\mathbf{n}) a(\mathbf{n}) + s_L \sum_{\braket{\mathbf{n'}\mathbf{n}}}a^\dagger(\mathbf{n'}) a(\mathbf{n'}),
\end{align}
with local smearing parameter $s_L$.
The local interaction is
\begin{align}
    V_L = \frac{q_L}{2}\sum_{\mathbf{n}} :\rho_L(\mathbf{n})\rho_L(\mathbf{n}):,
\end{align}
and the non-local interaction is
\begin{align}
    V_{NL} = \frac{q_{NL}}{2}\sum_{\mathbf{n}} :\rho_{NL}(\mathbf{n})\rho_{NL}(\mathbf{n}):.
\end{align}
The symbols :: indicate normal ordering, where the annihilation operators are on the right-hand side and the creation operators are on the left-hand side. The parameters $q_L, q_{NL}, s_L,$ and $s_{NL}$ are tuned so that both the interactions $H_A = K + V_L$ and $H_B = K + V_{NL}$ give the physical binding energy for ${}^{4}$He. The kinetic term is taken to be \cite{Epelbaum_2010},
\begin{align}
    K = \frac{49}{12m}\sum_{\mathbf{n}} a^\dagger(\mathbf{n}) a(\mathbf{n}) - \frac{3}{4m}\sum_{\mathbf{n},i}  \sum_{\braket{\mathbf{n'}\mathbf{n}}_i} a^\dagger(\mathbf{n}) a(\mathbf{n}) + \frac{3}{40m}\sum_{\mathbf{n},i}  \sum_{\braket{\braket{\mathbf{n'}\mathbf{n}}}_i} a^\dagger(\mathbf{n}) a(\mathbf{n}) - \frac{1}{180m}\sum_{\mathbf{n},i}  \sum_{\braket{\braket{\braket{\mathbf{n'}\mathbf{n}}}}_i} a^\dagger(\mathbf{n}) a(\mathbf{n}).
\end{align}

\subsection{Auxiliary field formalism}
In our Monte Carlo simulations, we use the auxiliary-field formalism, where we replace the nucleon-nucleon interactions with interactions of nucleons with background auxiliary fields at every lattice point. The main idea behind this method can be understood as a Gaussian integral formula,
\begin{align}
    :\exp \Big( -\frac{1}{2}ca_t\rho^2\Big): = \sqrt{\frac{1}{2\pi}} \int_{-\infty}^{+\infty} ds :\exp \Big( -\frac{1}{2}s^2 + \sqrt{-ca_t}s\rho \Big):.
\end{align}
We can reproduce the interactions by integrating over the auxiliary fields. For simulating two-, three- and four-body forces on the lattice simultaneously,  we can write the discrete auxiliary fields in the form,
\begin{align}
    :\exp \Bigg( -\frac{1}{2}Ca_t\rho^2 -\frac{1}{6}C_3a_t\rho^3 -\frac{1}{24}C_4a_t\rho^4 \Bigg): = \sum_{k=1}^N \omega_k :\exp\Big(\sqrt{-Ca_t}\phi_k\rho\Big):, \label{Eq:Aux_field_def}
\end{align}
where $C,C_3$, and $C_4$ are two-, three- and four-body coefficient respectively, and the :: symbols indicate normal ordering of operators. We use the ansatz $N = 3$, and $w_k>0$ for all $k$. We expand Eq.~\eqref{Eq:Aux_field_def} up to $O(\rho^4)$ and compare both sides order by order to solve for $\omega_k$ and $\phi_k$. For our case, we only have a two-body interaction ($C_3 = C_4 = 0$), and the solution is $\phi_1 =  -\phi_3 = \sqrt{3}, \omega_1 = \omega_3 = 1/6, \omega_2 = 2/3$. We then sample the corresponding auxiliary fields with the shuttle algorithm, as described in Ref.~\cite{Lu:2018bat}.

\subsection{Eigenvector continuation results for $^{12}$C}

\begin{figure}[hbt!]
    \includegraphics[width=12.2cm]{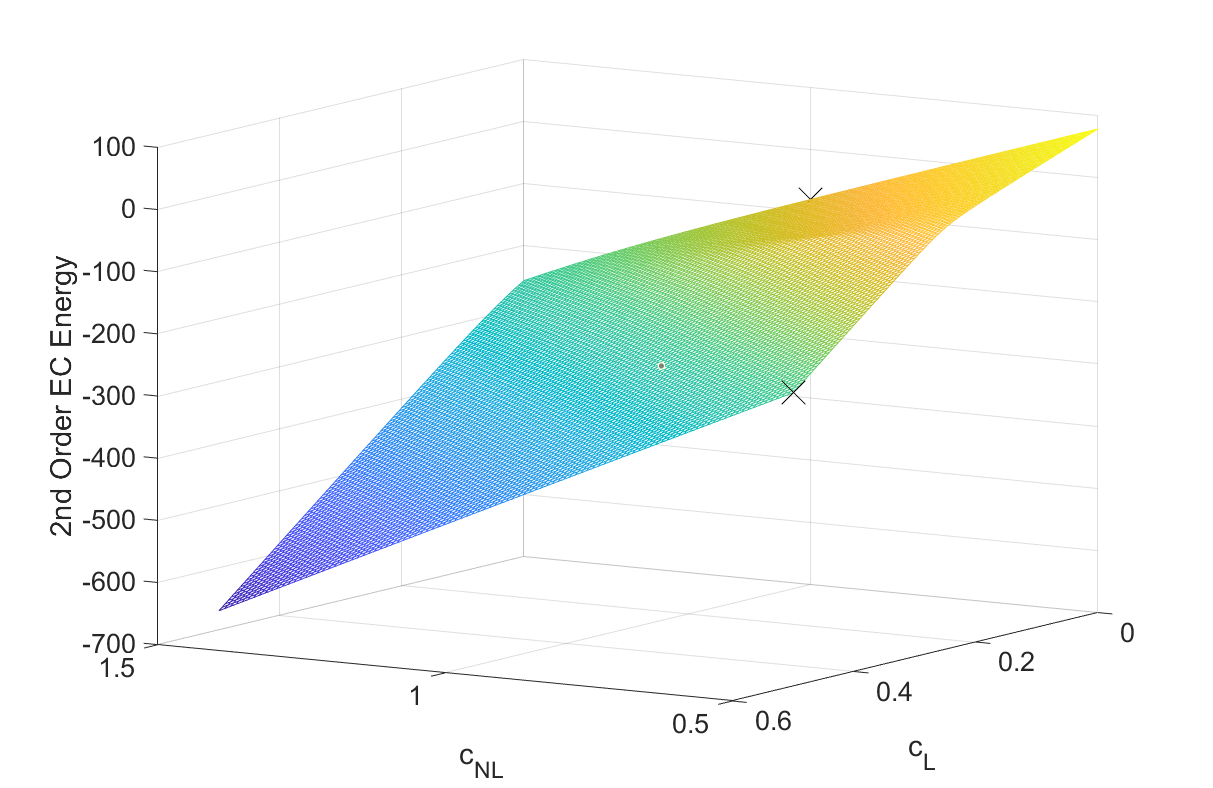}\\
    \caption{EC emulator results for ${}^{12}$C with two training points. The training points are at $(c_L,c_{NL}) = (0.5,0.5)$ and $(0,1)$, and are marked as X.
    }
    \label{Fig:E_C12_3d_plot}
\end{figure}

In Fig.~\ref{Fig:E_C12_3d_plot} we show the EC ground state energy for ${}^{12}$C with two training points chosen at $(c_L,c_{NL}) = (0.5,0.5)$ and $(0,1)$, with $L_t = 600$. There is a sharp bend in the slope of the ground-state energy, and similar results are seen for the simulations of ${}^{8}$Be and ${}^{16}$O.  This sharp bend in the slope of the energy coincides with the location of the quantum phase transition between a Bose gas of alpha particles and a nuclear liquid. 

\subsection{Computational advantage of auxiliary field reordering}
The computational advantage we achieve by reordering auxiliary fields in the floating block method depends on the number of Euclidean time steps, $L_t$, and the choice of training points. If the training eigenvectors are very similar to each other and $L_t$ is not too large, then the advantage of auxiliary field reordering is relatively minor. If the training eigenvectors are different, however, the computational advantage of reordering is substantial.  If we consider the ${}^{12}$C system with training points $(c_L,c_{NL}) = (0.5,0.5)$ and $(0,1)$ as presented in the main text, we find that the Monte Carlo results are extremely noisy if we do not reorder the auxiliary fields in the numerator of Eq.~\eqref{N_ij}. For large values of $L_t$, the calculations without auxiliary field reordering are so poor that the relative errors exceed $100\%$ and fluctuate wildly.  For a quantitative benchmark of the computational advantage provided by auxiliary field reordering, it is therefore useful to choose a calculation with a small number of time steps.  For this purpose, we consider the squared norm matrix element $|N_{12}|^2$ for ${}^{12}$C with $L_t = 4$ and compare results obtained with and without auxiliary field reordering.  To make a fair comparison, we keep the number of Monte Carlo trials the same for both calculations. When we reorder the auxiliary fields, we get $|N_{12}|^2 = 0.999989(2)$.  When we don't reorder the auxiliary fields, we get $|N_{12}|^2 = 1.15(20)$.  We see that the reduction in error is five orders of magnitude for this test calculation.  An error reduction by five orders of magnitude is equivalent to increasing the number of trials by ten orders of magnitude.  This benchmark test shows that the computational advantage of auxiliary field reordering is many orders of magnitude for this example and should grow exponentially with system size and Euclidean time.  For the calculations presented in Table~\ref{Table:exact_error}, we simply cannot calculate norm matrix elements without reordering auxiliary fields.  

\subsection{Quantum Monte Carlo reweighting calculations}
In previous work such as Ref.~\cite{Frame:2019jsw}, norm matrix elements and Hamiltonian matrix elements are calculated using reweighting.  This consists of sampling auxiliary field configurations according to some reference Hamiltonian coupling $c_k$. The calculations to be performed are then
\begin{align}
    H_{ij}(c_t) & \propto \frac{\braket{\psi_{I}|c_i|\cdots|c_i|H(c_t)|c_j|\cdots|c_j|\psi_{I}}}{\braket{\psi_{I}|c_k|\cdots|c_k|c_k|\cdots|c_k|\psi_{I}}},\\
    N_{ij} & \propto \frac{\braket{\psi_{I}|c_i|\cdots|c_i|c_j|\cdots|c_j|\psi_{I}}}{\braket{\psi_{I}|c_k|\cdots|c_k|c_k|\cdots|c_k|\psi_{I}}}.
\end{align}
We can then calculate the Hamiltn and norm matrix elements up to an irrelevant overall factor.  For large systems for large $L_t$, the Monte Carlo importance sampling will be poor if the ground state wave functions for $H(c_i)$ or $H(c_j)$ are significantly different from the ground state wave function for $H(c_k)$.  When the ground state energies $E(c_i)$ and $E(c_j)$ are different from $E(c_k)$, the importance sampling problem becomes even worse due to exponential factors of $\exp\{L_t[E(c_k)-E(c_i)]/2\}$ and $\exp\{L_t[E(c_k)-E(c_j)]/2\}$ appearing in the ratios.  As a result, the reweighting approach is viable only when the system is small in size, the Euclidean time $t$ is not too large, and the ground state energies for the different Hamiltonians are numerically close.

\subsection{Computational advantage of the floating block method}
In this section, we do a benchmark comparison of the floating block method versus the reweighting approach described in the previous section.  For the reweighting calculations, let us define the reweighting ratios
\begin{align}
    N'_{ij} & = \frac{\braket{\psi_{I}|c_i|\cdots|c_i|c_j|\cdots|c_j|\psi_{I}}}{\braket{\psi_{I}|c_j|\cdots|c_j|c_j|\cdots|c_j|\psi_{I}}},\\
    N'_{ii} & = \frac{\braket{\psi_{I}|c_i|\cdots|c_i|c_i|\cdots|c_i|\psi_{I}}}{\braket{\psi_{I}|c_j|\cdots|c_j|c_j|\cdots|c_j|\psi_{I}}}.
\end{align}
We note that
\begin{align}
    \frac{|N'_{ij}|^2}{N'_{ii}} = \frac{|\braket{\psi_{I}|c_i|\cdots|c_i|c_j|\cdots|c_j|\psi_{I}}|^2}{\braket{\psi_{I}|c_i|\cdots|c_i|c_i|\cdots|c_i|\psi_{I}} \braket{\psi_{I}|c_j|\cdots|c_j|c_j|\cdots|c_j|\psi_{I}}}.
\end{align}
In the limit of large $L_t$, this equals the squared norm matrix element $|N_{ij}|^2$,
\begin{equation}
     \frac{|N'_{ij}|^2}{N'_{ii}} = |N_{ij}|^2.
\end{equation}
We can now directly compare the computational performance of the reweighting technique for ${|N'_{ij}|^2}/{N'_{ii}}$ and the floating block method for $|N_{ij}|^2$.

We perform the reweighting and floating block calculations for the $^{12}$C system for a box size with length $L=15.76$ fm.  We calculate $|N_{12}|^2$, where the two training points are chosen to be $(c_L,c_{NL}) = (0.5,0.5)$, and $(0.0,1.0)$. The results for $L_t = 200$ and $L_t = 300$ time steps are shown in Table~\ref{Table:FB_vs_noFB}.

\begin{table*}[hbt!]
\centering{}
\begin{tabular}{| c| c| c|}
\hline
 $L_t$ & 200 & 300 \\ 
 \hline
 $|N_{12}|^2$ (using floating block) & $0.504 \pm 0.011$ & $0.203 \pm 0.011$ \\ 
 ${|N'_{ij}|^2}/{N'_{ii}}$ (using reweighting) & $1.5 \pm 1.0 \times 10^5$ & $0.8 \pm 1.1 \times 10^6$ \\ 
\hline
\end{tabular}
\caption{Comparison of norm matrix calculations for ${}^{12}$C calculated using the floating block and reweighting approaches. The two training points are at $(c_L,c_{NL}) = (0.5,0.5)$, and $(0.0,1.0)$.
}
\label{Table:FB_vs_noFB}
\end{table*}

Our results show that the reweighting technique has errors that are seven orders of magnitude larger than that of the floating block method for $L_t = 200$ and eight orders of magnitude larger for $L_t = 300$.  The floating block method is clearly providing a computational advantage that is equal to many orders of magnitude for this example and should grow exponentially with system size and Euclidean time. The reweighting approach is not able to perform any of the calculations presented in the main text.



\end{document}